# iDART-Intruder Detection and Alert in Real Time

Manish Kumar, Shubham Kaul, Vibhutesh Kumar Singh, Vivek Ashok Bohara
{manish12161, shubham12161, vibhutesh1494, vivek.b}@iiitd.ac.in
Wirocomm Research Group, Indraprastha Institute Of Information Technology (IIIT Delhi),
New Delhi

**Abstract—** In this work, we design and develop a smart intruder detection and alert system which aims to elevate the security as well as the likelihood of true positive identification of trespassers and intruders as compared to other commonly deployed electronic security systems. Using multiple sensors, this system can gauge the extent of danger exhibited by a person or animal in or around the home premises, and can forward various critical information regarding the event to home owners as well as other specified entities, such as relevant security authorities.

## I. INTRODUCTION

Various security systems can be found installed in homes and offices alike. For example, Closed-circuit Television (CCTV) is a popular technology used extensively for the purpose of home & office security and surveillance. Although, such systems may offer some merits in the form of low cost and relative ease of installation, however they also have some major drawbacks. These systems usually have no mechanism to send critical information to users/home owners if they are not present in the home premises. Additionally, since the current systems record video continuously and store it locally, one has to scroll through long duration of elapsed time often hours, in order to get to the relevant section of the video. The system presented in this paper aims to improve upon such shortcomings of other security systems. This system sends relevant video data (on detection of an intrusion) to the users and home owners directly, without the need of an additional centralized monitoring unit. Since all of this is done in real time, this data could also be relayed to the security authorities who would expedite the process of identification or capturing of an intruder.

## II. IMPLEMENTATION and KEY FEATURES

By using a combination of sensors and microprocessors that are described in the following sections, this system is able to incorporate some crucial features into its functioning. These are explained below:

This system uses smart sensors to monitor the presence of trespassers and possible intruders near the homes entry points such as door, windows etc. This feature enables the system to start recording video only when it perceives the danger. Hence, a user does not have to scan through hours of irrelevant video data[6]. This not only saves time but also decreases the overall memory and processing power requirements.

Furthermore, the implemented system uses the locally available Wireless LAN (WLAN) [3] network to send an e-mail to the user with the video recorded as an attachment. Thus no additional hardware installations are required to alert the user.

This system also uses a sensor placed on the inside of the secured premises entry points, i.e., near doors/windows etc in order to detect intrusions and send mail immediately to the home owner as well as security authorities in a case of intrusion. We have depicted three test cases in which our system works, which are illustrated through Figure 1, Figure 2 & Figure 3.

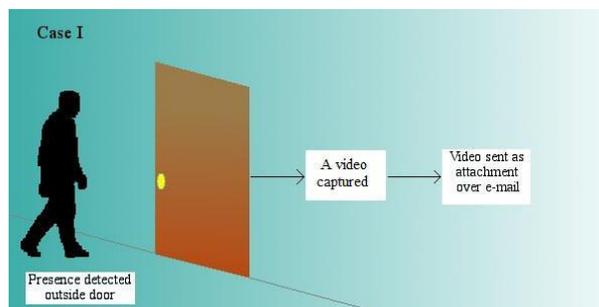

Figure 1. System response when presence detected near door

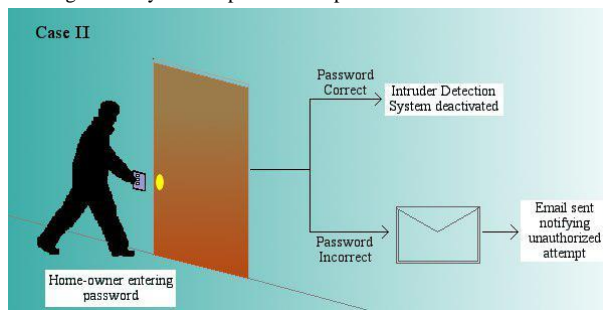

Figure 2. System response when password mechanism is used

Since this system works on wireless protocols such as IEEE 802.11 [3], ZWave [5] and Zigbee Protocol [1] (explained in later sections), it could also be extended to interact with other devices very easily, especially with the already installed Home Automation System (HAS) working with same wireless protocol, like [2], in order to maximize security & reliability.

Since in the event of a friendly visit, the home owner or other authorized member would not want to trigger the security alarm system, there is also an integrated deactivating mechanism placed outside the door, implemented in the system, with an unique PIN based password input system.

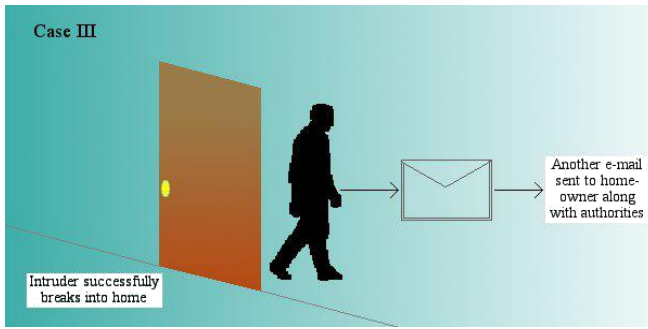

Fig. 3. System response in case of a break-in

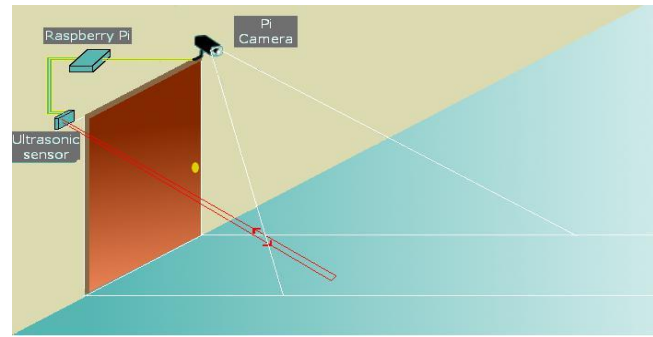

Figure. 5. Hardware Setup outside the house.

It also presents an alternative password input mechanism to traditional PIN-based password systems (explained in section III).

## III. HARDWARE SETUP

To implement the system with mentioned features, three processing units have been used- a microcontroller, a raspberry pi [5] single board computer and a PC (with LabVIEW installed). This approach made our system modular since if we need to change any feature, we only require to change that particular module not the whole setup, moreover our system is portable enough to occupy less space while installation.

The microcontroller unit wirelessly sends an intruder alert to the central PC in case of a break-in, via Zigbee protocol [1]. To accomplish this, the microcontroller uses a combination of a Laser and a LDR pair as an input, to detect an unauthorized access.

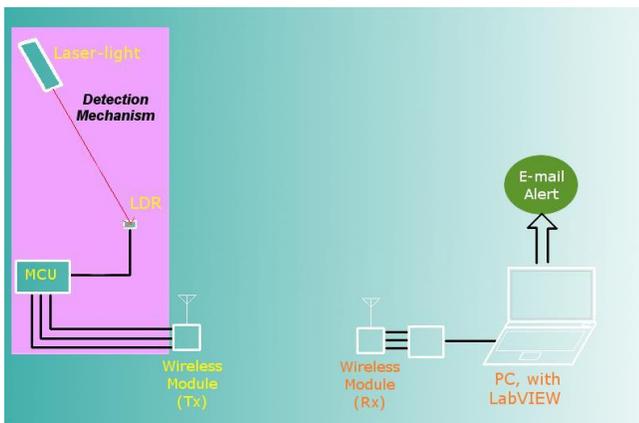

Figure. 4. Hardware Setup inside the house.

Figure 4 depicts a schematic of our implemented system, when installed inside the premise under observation. Once activated, any kind of obstruction of the laser beam used in our prototype triggers the wireless module attached with microcontroller to send the intruder alert. The deactivating mechanism for this system is also controlled by the microcontroller unit. The input for the deactivation mechanism is placed intelligently outside the door and only an authorized person will be able to insert the password and activate/deactivate the system.

The central gateway computer, after receiving an intruder alert from the microcontroller unit sends an email using the LABVIEW based software, to the concerned user as well as security agencies regarding the intrusion.

The Raspberry Pi Single Board Computer on the other hand handles the image/video capturing aspect of the project, as it requires more memory & processing power. The Raspberry Pi single board computer uses an ultrasonic sensor to detect the presence of an individual near the door. Once the sensor is triggered by the presence of a person, it starts recording a 5-10 seconds video of that area, which capture the identity of the trespasser or any activity happening outside the door on footage. This video is then both saved on-board and sent to the home owner and/or authorities via e-mail.

## IV. FUTURE WORK

Further, this project can be easily expanded to include various additional features, functionalities and services. For example, the password/deactivation mechanism could be replaced from the current button-input style to a traditional PIN based system or even replaced with a voice-command based activation/deactivation for certain part of this security system. Also, in order to completely monitor all activity near the home premises, multiple Raspberry Pi based units can be integrated instead of just one. Such Raspberry Pi computer units can be located at all possible entry points such as windows, vents, etc., and not just the main door. Since the Raspberry Pi single board computers operate independently of the microcontroller unit in our implementation, such expansion would be extremely easy. Finally, the Zigbee protocol is immensely used for inter-device communication and IOT implementations. Hence, this provides endless possibilities in which this system could be used in conjunction with other Zigbee enabled devices. For cases where the home owner is connected to the internet, he/she may not be able to check the emails sent by this security system. To solve this issue, an SMS sending feature can also be integrated in the current system easily, which will immediately notify the owner and concerned authorities about the intrusion/unauthorized access.